\documentclass[reprint,aps,prb,floatfix,longbibliography,superscriptaddress]{revtex4-2}

\usepackage{graphicx}
\usepackage[caption=false,labelformat=empty]{subfig}
\usepackage{amsmath,amssymb}
\usepackage{booktabs}
\usepackage{microtype}            
\usepackage[colorlinks=true,allcolors=blue]{hyperref}
\usepackage{bm}
\begin{document}

\preprint{APS/123-QED}
\title{Even-harmonic generation from topological edge states in generalized Su-Schrieffer-Heeger models}

\author{Chi-Ting Liu}
\affiliation{Graduate Institute of Photonics and Optoelectronics and Department of Electrical Engineering, National Taiwan University, Taipei 10617, Taiwan}
\affiliation{Graduate Institute of Applied Physics, National Chengchi University, Taipei 11605, Taiwan}
\author{J-S You}
\email{jhihshihyou@ntnu.edu.tw}
\affiliation{Department of Physics, National Taiwan Normal University, Taipei 11677, Taiwan}

\author{Hsiu-Chuan Hsu}
\email{hcjhsu@nccu.edu.tw}
\affiliation{Graduate Institute of Applied Physics, National Chengchi University, Taipei 11605, Taiwan}
\affiliation{Department of Computer Science, National Chengchi University, Taipei 11605, Taiwan}

\date{\today}

\begin{abstract}
High-order harmonic generation (HHG) in solids has emerged as a powerful probe of symmetry and topological properties in quantum materials. In this work, we investigate the HHG response in one-dimensional solids with edge or midgap states under global and local illumination. We numerically compute the HHG spectrum for the Su-Schrieffer-Heeger (SSH) model with next-nearest-opposite sublattice hopping, dubbed the extended SSH (ESSH) model, and the Rice-Mele model, a one-dimensional system with broken inversion symmetry introduced via staggered on-site potentials. By contrasting the spectral features of the ESSH and Rice-Mele models under global illumination, our analysis reveals that although midgap states provide additional pathways for transitions, the resulting interference is destructive, leading to spectral features distinct from those of edge states. Furthermore, when a single boundary of the topological insulator is locally illuminated, the HHG spectrum of the edge states exhibits vanishing odd harmonics, leaving even harmonics dominant in the spectrum. We identify this even-harmonic selection rule as a consequence of the zero-energy character of the edge states and the particle-hole symmetry of the system, which enforces even field parity of the zero-mode response. These findings reveal that the spatial location of the laser illumination offers a route to control the symmetry of the system, thereby selectively suppressing or enhancing even- and odd-order harmonics in low-dimensional nanostructures.
\end{abstract}

\maketitle

\section{\label{sec:introduction} INTRODUCTION}
High-order harmonic generation (HHG) is a hallmark of nonlinear optics in which radiation at multiples of a driving field's frequency is emitted when the target system interacts with an intense ultrashort laser pulse. Initially observed in gaseous atoms and molecules~\cite{PhysRevLett.70.774, PhysRevLett.68.3535,McPherson:87,Hu2017}, the process is well described by the semiclassical three-step model: (i) tunnel ionization, (ii) field-driven propagation of the freed electron, and (iii) recollision with the parent ion, leading to high-energy photon emission. This framework has been remarkably successful in elucidating the fundamental mechanisms of HHG in atomic systems. With the seminal study, HHG has been demonstrated to be applicable to solid-state systems like bulk crystal and semiconductors~\cite{Ghimire2019,Observation_of_HHG_in_a_bulk_crystal, Real-time_observation_of_interfering_crystal_electrons_in_HHG, Linking_HHG_from_gases_and_solids, Merge_of_HHG_from_gases_and_solids_and_its_implications_for_attosecond_science,PhysRevLett.113.073901,PhysRevLett.113.213901,PhysRevLett.120.253201,OPTICA.402393,Vampa2016}, creating novel avenues for ultrafast spectroscopy and the probing of complex electronic dynamics in solids. In contrast to atomic systems, solids exhibit a rich band structure, necessitating a solid-state generalization of the three-step model that accounts for both intraband (electron motion within a band) and interband (electron motion between two bands) transitions of Bloch electrons and holes.

More recently, HHG has been drawing growing interest in the field of condensed-matter as an all-optical method for investigating topological phases. In particular, Bauer and Hansen~\cite{PhysRevLett.120.177401} demonstrated that the HHG spectrum for topological and trivial phases can exhibit tremendous differences within the bulk bandgap regime by using time-dependent density-functional theory (TDDFT)~\cite{PhysRevLett.52.997}. Complementing these results, Jürß  and Bauer~\cite{PhysRevB.99.195428} reproduced similar features using the Su-Schrieffer-Heeger (SSH) model \cite{Su1979}, showcasing the utility of simple tight-binding approaches in capturing topological signatures in HHG spectrum. Chuan Yu \cite{PhysRevA.110.033113} further explained how the topological edge states (ESs) enhance the harmonic emission and the mechanisms of the interference between quantum pathways by implementing solid-state three-step model. Notably, HHG was shown to be sensitive not only to the presence but also to the number of ESs \cite{PhysRevB.108.214104}, enabling differentiation between phases with zero, two, or four edge states in the extended SSH (ESSH) model. Beyond the SSH-type systems, Liu and Bian~\cite{PhysRevB.105.054308} extended this work to the topologically nontrivial Aubry-André-Harper(AAH) model~\cite{PhysRevLett.109.116404, PhysRevB.91.014108}. By modulating the phase $\phi$ to alter the hopping amplitude, they also demonstrated that ESs can be effectively detected via enhancement of HHG below the bandgap, with strong robustness in both periodic and quasiperiodic systems. 

Several studies have paid attention to the role of symmetry-breaking terms. Ma \textit{et al.}~\cite{PhysRevB.106.125117} addressed this concern by considering open-boundary SSH chains via Lindblad master equation to incorporate quantum decoherence. They studied HHG from a selected single-edge state in an open SSH chain and reported even-order harmonic emission.  They further analyzed how symmetry breaking and decoherence modify the odd-even harmonic structure using a two-channel interference model and dynamical symmetry arguments. Thereafter, Nivash \textit{et al.}~\cite{PhysRevB.110.115103} introduced an on-site potential of the cosine wave form in the AAH model. It can be effectively reduced to a staggered potential, analogous to a staggered on-site potential in the SSH model. Their results showed that the on-site potential would weaken the harmonic spectrum in the bulk bandgap regime, and the harmonic spectrum intensity is strongly dependent on the strength of the on-site potential. 

In this work, we explore the HHG response in the extended SSH model and Rice-Mele model \cite{PhysRevLett.49.1455, PhysRevB.102.085122}. 
With the extended coupling, both models enrich the band structure and provide additional quantum pathways for electrons. As the edge states and midgap states are characteristics of the SSH and RM model, respectively, we study how the edge states in the extended SSH model and the midgap states in the Rice-Mele model affect the interference and reshape the spectral profile. Furthermore, we investigate the harmonic spectra of edge states by local illumination and perform the symmetry analysis. The results would shed light on the lightly doped semiconductors and localized impurity states. 
For clarity, we use the abbreviation ESs to denote the edge states in both the SSH and extended SSH models, and MGSs to denote the midgap states in the Rice-Mele model.

This paper is organized as follows. To begin with, the theoretical method and model used in this study are presented in Sec.~\ref{sec:method}, including extended SSH and Rice-Mele model, the coupling field and the numerical method for the time evolution of the wavefunction. The calculated harmonic spectra are given in Sec.~\ref{sec:results}.  The symmetry analysis for the emergence of the even harmonics from the edge states is presented in Sec.~\ref{sec: local_illuminating}. 
Finally, this paper is summarized in Sec.~\ref{sec: summary}. 

\section{\label{sec:method} METHOD}
\subsection{\label{subsec:ESSH and Rice-Mele model} Generalized Su-Schrieffer-Heeger models and Rice-Mele Model}
\begin{figure}[tbp]
	\includegraphics[width=\linewidth]{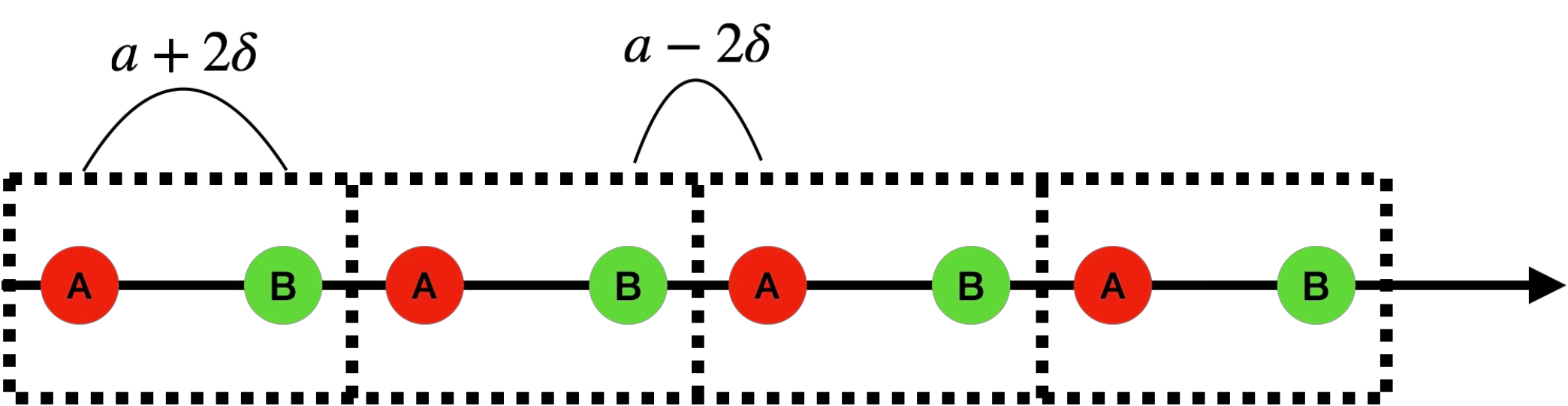}
	\caption{The schematic of the one-dimensional chain studied in this work. Each dashed rectangle denotes a unit cell containing two sublattices A and B. The distance between A and B within one unit cell is $a+2\delta$. The distance between A and B of neighboring unit cells is $a-2\delta$.}
	\label{fig:lattice}
	\end{figure}

Topological insulators~\cite{RevModPhys.83.1057, Batra2020,RevModPhys.82.3045,doi:10.7566/JPSJ.82.102001} are characterized by topological invariants and constitute a distinct class of quantum materials: they are insulating in the bulk while hosting conducting boundary states. The one-dimensional Su-Schrieffer-Heeger (SSH) model is a paradigmatic tight-binding framework widely employed to study topological phases in crystalline systems. It describes a dimerized chain with two sublattices, $A$ and $B$, arranged in a primitive unit cell, where the hopping integrals between nearest-neighbor sites include intracell hopping amplitude $v$ and intercell hopping amplitude $w$. The schematic of the one-dimensional lattice is shown in Fig. \ref{fig:lattice}. The one dimensional field free Hamiltonian of the SSH model is given by:
\begin{equation}
\begin{aligned}
H_{SSH} &= v \sum_{n=1}^{N/2} \hat{c}_{n,a}^{\dagger} \hat{c}_{n,b} 
&+ w \sum_{n=1}^{N/2-1} \hat{c}_{n,b}^{\dagger} \hat{c}_{n+1,a} + \text{H.c.}
\end{aligned}  
\end{equation}
where $N$ is the total number of sites, $n$ indexes the unit cells, $\hat{c}_{n,a}$ ($\hat{c}_{n,a}^{\dagger}$) and $\hat{c}_{n,b}$ ($\hat{c}_{n,b}^{\dagger}$) are annihilation (creation) operators for sublattices $A$ and $B$, respectively. The relative strength of $v$ and $w$ determines the topological phase: for $|v| > |w|$ the system is a trivial insulator, while for $|v| < |w|$ it enters a topological phase featuring zero-energy edge states and integer winding number.

To explore more intricate topological properties, we extend the SSH model by including next-nearest-neighbor (NNN) hopping, leading to the extended SSH (ESSH) model 
\begin{equation}
	\begin{aligned}
		H_{ESSH} &=H_{SSH}+ J \sum_{n=1}^{N/2-2} (\hat{c}_{n,b}^{\dagger} \hat{c}_{n+2,a} + \text{H.c.}),
\end{aligned}  
\end{equation}	
where $J$ is the long-range hopping between A,B sublattices of next-nearest-neighboring unit cells. This extension enables the exploration of enriched topological features and allows the presence of additional zero-energy edge states. The SSH and ESSH models preserve inversion symmetry, chiral symmetry ($\Gamma$), spinless time-reversal symmetry $K$ (complex conjugation), and particle-hole symmetry ($\mathcal{P}$). In the single-particle sector, the chiral operator is represented as $\Gamma = \sum_{n}(\hat{c}^{\dagger}_{n,a}\hat{c}_{n,a} -  \hat{c}^{\dagger}_{n,b}\hat{c}_{n,b})$, with eigenvalues $+1$ and $-1$ on the A and B sublattices, respectively. Following the general relation $\Gamma=\mathcal{P}{K}$ of the Altland--Zirnbauer classification~\cite{Altland1997,Schnyder2008,Chiu2016}, the particle-hole operator is given by $\mathcal{P}=\Gamma K$ for these models.

Additionally, to examine inversion-symmetry breaking, we also consider the Rice-Mele model,  
\begin{equation}
\begin{aligned}
	H_{RM}&=H_{ESSH}+H_U\\
H_U &= U\sum_{n=1}^{N/2} (\hat{c}_{n,a}^{\dagger} \hat{c}_{n,a} - \hat{c}_{n,b}^{\dagger} \hat{c}_{n,b}).
\end{aligned}
\end{equation}
where $H_U$  is the staggered on-site potential, of which magnitude is denoted by $U$.

\begin{figure*}[tbp]
    \centering
    \includegraphics[width=0.98\linewidth]{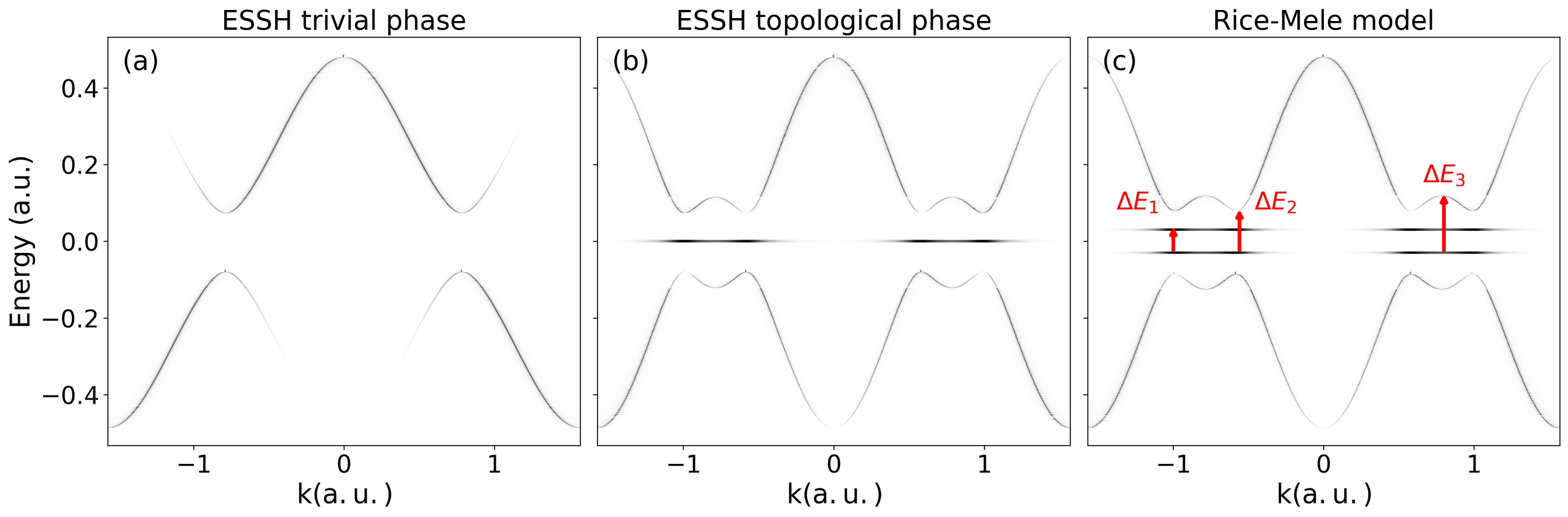}
  \caption{Momentum-resolved energy spectra of the ESSH-trivial (a), ESSH-topological (b) and the Rice-Mele models (c) for $N=2000$. Red arrows in (c) indicate three energy differences defined in Table \ref{tab:param}: $\Delta E_1 \approx 0.06$, $\Delta E_2 \approx 0.11$, and $\Delta E_3 \approx 0.15$.}
  \label{fig:bandstructures}
\end{figure*}

In this paper, we study the HHG spectrum for three representative cases: ESSH-topological phase, ESSH-trivial phase and Rice-Mele (RM) model. The parameters are chosen to ensure that the three models have similar energy gaps and total spectral widths for a better comparison. The atomic unit (a.u.) $\hbar = |e| = m_{\mathrm{e}} = 4 \pi \varepsilon_0 = 1$ are used throughout this study. Throughout this paper, we set $v=-0.25, w=-0.20, J=-0.03$ for ESSH-trivial and $v=-0.10, w=-0.18, J=-0.2$ for both  ESSH-topological and RM. The corresponding winding number for the ESSH-topological model is $2$, suggesting $4$ edge states. Momentum-resolved energy spectra for the ESSH-trivial phase, ESSH-topological phase, and the RM model with the chosen parameters are shown in Fig.~\ref{fig:bandstructures}. The bulk energy gaps ($E_{gap}$) of the ESSH-trivial and ESSH-topological models are $\approx 0.15$ and that of the RM model is $\approx 0.16$. The total spectral widths ($E_{max}$) for three models are $\approx 0.97$. The ESSH-topological phase hosts four near-zero edge states, whose spectral weight appears around zero energy in Fig.~\ref{fig:bandstructures}(b). 
For the RM model, as shown in Fig.~\ref{fig:bandstructures}(c), due to the on-site potential, the original four degenerate edge states are split into two groups of midgap states: one pair shifts upward in energy by $+U$, and the other shifts downward in energy by $-U$ relative to zero energy. The characteristic energy differences for the energy bands in  Fig.~\ref{fig:bandstructures} are summarized in Table \ref{tab:param} 

\begin{table}[]
	\begin{tabular}{|c|c|c|}
		\hline
		\begin{tabular}[c]{@{}c@{}}Energy level \\ differences\end{tabular} & Symbol       & \begin{tabular}[c]{@{}c@{}}Corresponding \\ harmonic order$^{\dagger}$\end{tabular}    \\ \hline
		Total spectral width$^*$                                            & $E_{max}$    & $\approx 129$                                                                          \\ \hline
		Bulk band gap                                                       & $E_{gap}$    & \begin{tabular}[c]{@{}c@{}}$\approx 20$ (ESSH)\\ $\approx 22$ (Rice-Mele)\end{tabular} \\ \hline
		Midgap                                                      & $\Delta E_1$ & $8$                                                                                    \\ \hline
		CB minimum - lower MGS                                              & $\Delta E_2$ & $15$                                                                                   \\ \hline
		CB local maximum - lower MGS                                        & $\Delta E_3$ & $\approx 20$                                                                           \\ \hline
	\end{tabular}
\caption{ The table summarizes the characteristic energy differences used in the paper. 
	$^*$ Total spectral width refers to the maximum energy difference between conduction and valence band. $^{\dagger}$ The corresponding harmonic orders are calculated using the model parameters for Fig. ~\ref{fig:bandstructures} and pulse frequency $\omega=0.0075$ a.u. }
\label{tab:param}
\end{table}

\subsection{\label{subsec: harmonic spectrum} Harmonic spectrum}
To investigate high-order harmonic generation, we couple the system to a linearly polarized laser pulse along $x$-axis, $\mathbf{E}(t)=E(t)\hat{x}$,  via a time-dependent vector potential field within the dipole approximation. The electric field is computed as the time derivative of the vector potential
\begin{equation}
	{E}(t) = -\partial_t A(t), 
\end{equation}
where $A(t)$ is simulated as 
\begin{equation}
    A(t) = A_0 \sin^2 \left( \frac{\omega t}{2 n_{\mathrm{cyc}}} \right) \sin \omega t, \quad 0 < t < \frac{2\pi n_{\mathrm{cyc}}}{\omega}.
\end{equation}
Outside the pulse duration, $A(t)$ is set to zero. This specific sin-squared envelope ensures a smooth turn-on and turn-off of the field, avoiding abrupt changing that could introduce numerical instabilities. 
In our calculations, we adopted length gauge to simulate the coupling to laser fields. For global illumination, we added $\sum_{j=1}^NE(t)x_j$ in the full Hamiltonian $\mathcal{H}$. For local illumination, 
the electric field becomes not only time dependent but also position dependent which takes the form \cite{PhysRevB.99.195428}:
\begin{equation}
E(x, t) = E(t) \cos^2\left( \frac{(x - x_0)\pi}{2 x_l} \right)
\end{equation}
for $|x-x_0|<x_{\ell}$ and $0$ otherwise, where $x_0$ denotes the center of the illumination region, and $x_l$ defines the spatial extent of the pulse. In the length gauge, the potential energy is added to the onsite terms of the Hamiltonian 

\begin{equation}
    \epsilon_j = d_j E(t).
   \label{eq:localpot}
\end{equation} 
where $d_j=\int_0^{x_j}\cos^2\left( \frac{(x - x_0)\pi}{2 x_l}\right)dx$.
The magnitude of the vector potential is set to $A_0 = 0.2\ \mathrm{a.u.}$, corresponding to a peak intensity of approximately $7.9\times 10^{10}\,\mathrm{W\,cm^{-2}}$, the fundamental frequency of the pulse $\omega$ is set to $0.0075\ \mathrm{a.u.}$ This long wavelength ensures the validity of the dipole approximation, assuming the spatial extent of the model system is much smaller than  wavelength $\lambda$. The number of optical cycles is fixed at $n_{\mathrm{cyc}} = 5$, generating a multi-cycle pulse suitable for frequency-domain analysis. 

Furthermore, time propagation of the wavefunction is performed using the Crank-Nicolson method:
\begin{equation}
\begin{aligned}
|\psi_n(t + \Delta t)\rangle 
\approx 
\left[ \frac{1 - \mathrm{i}\frac{\Delta t}{2}\, \mathcal{H}\left(t + \frac{\Delta t}{2} \right)}{1 + \mathrm{i}\frac{\Delta t}{2}\, \mathcal{H}\left(t + \frac{\Delta t}{2} \right)} \right] |\psi_n(t)\rangle,
\end{aligned}
\end{equation}
where $\psi_n(t)$ for $n = 1,2,3 \dots, N/2$ is the time-evolution of the $n$th eigenstate. For half-occupancy, the maximum $n$ is $N/2$.  
The expectation value of position operator is calculated by:

\begin{equation}
X(t) = \sum_{n=1}^{N/2}  \langle\psi_n(t)| \hat{x} |\psi_n(t)\rangle,
\end{equation}
where $\hat{x}=\text{diag}(x_1,x_2...)$ is the position operator and $x_j$ is the position of $j$-th atom, following Refs.~\cite{PhysRevLett.120.177401,PhysRevB.99.195428}:
\begin{equation}
x_j = (j- \frac{N+1}{2})a - (-1)^{j}\delta
\end{equation}
for $j = 1,2,3,\dots,N$. In the numerical calculation, $a$ is set to be $2.0$ a.u., $\delta$ is set to be $0.15$ a.u. for ESSH-trivial phase and $-0.15$ a.u. for ESSH-topological phase and RM model. The harmonic spectrum is obtained by taking the modulus squared of Fourier transform of the dipole acceleration, 
\begin{equation}
S(\omega) \propto \bigl|\operatorname{FFT}\{W(t)\ddot X(t)\}\bigr|^{2}.
\end{equation}
FFT stands for fast Fourier transform and $W(t)$ is a window function; here the Hanning function is used to reduce spectral leakage. $\ddot{X}(t)$ is acceleration of the dipole which can be simply calculated by second time derivative of position operator $X(t)$.

\subsection{\label{subsec: current_decomposition_analysis} Decomposition analysis}

To elucidate the origin of the major contributions to the HHG response, we analyze the HHG spectrum by decomposing the total dipole acceleration into three components: contributions from all half-occupied states ($S_{\mathrm{occ}}$), from the valence band ($S_{\mathrm{VB}}$), and from the edge states ($S_{\mathrm{ESs}}$) or midgap states ($S_{\mathrm{MGSs}}$), obtained by the Fourier transform of the  
dipole acceleration projected onto VB, MG and ES
\begin{eqnarray}
	\ddot{X}_{VB}(t)&=&\frac{d^2}{dt^2}\sum_{n=1}^{N/2-2}\langle \psi_n(t)| \hat{x}|\psi_n(t)\rangle \\
	\ddot{X}_{MG/ES}(t)&=&\frac{d^2}{dt^2}\sum_{n=N/2-1}^{N/2}\langle \psi_n(t)| \hat{x}|\psi_n(t)\rangle.
\end{eqnarray}
It is important to note that due to the finite size of the system and numerical limitations, the edge states in the ESSH model are not exactly degenerate at zero energy. Instead, four near-degenerate edge states locate around zero energy, with eigenvalue magnitudes in the order of $10^{-18}$ to $10^{-16}$. To respect the inversion symmetry of the model, we assign alternating parity (odd and even) to these four states, ordered from lowest to highest in energy.

\section{\label{sec:results} HHG spectra of global and local illumination}
\begin{figure}[tbp]
    \centering
    \includegraphics[width=0.97\linewidth]{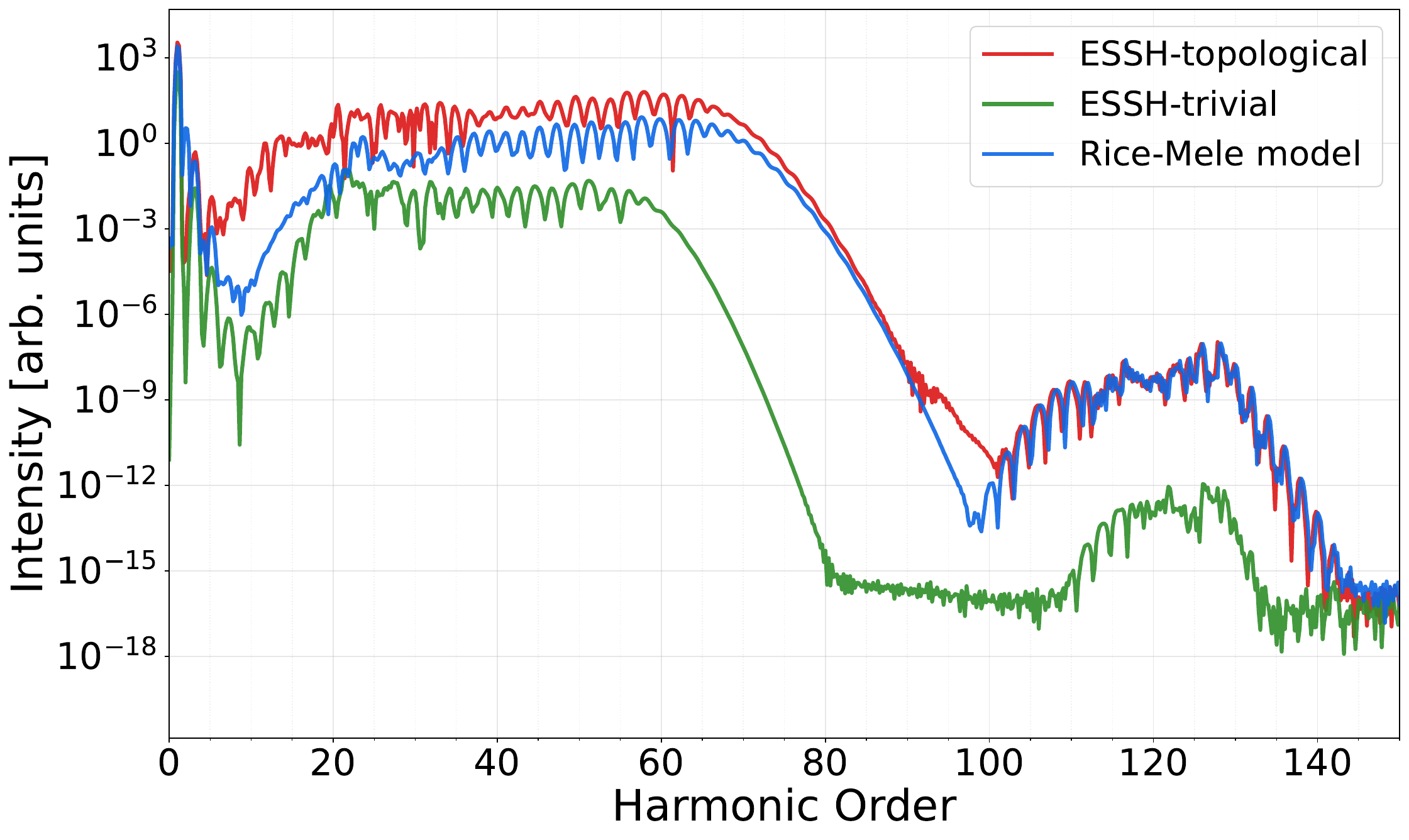}
    \caption{Harmonic spectra for the ESSH-trivial phase, ESSH-topological phase, and the Rice-Mele model for $N = 2000$, $a = 2$. 
    	 The blue and red dashed vertical lines denote harmonic orders of $8$ and $20.4$, respectively. }
    \label{fig:HHG_compare_essh_rice_mele}
\end{figure}

First, we present harmonic spectra of the models under global illumination. Fig. \ref{fig:HHG_compare_essh_rice_mele} presents the harmonic spectra of the ESSH-topological, ESSH-trivial and RM models under global illumination. 
The spectra can be understood with the three-step model which describes the process: (i) excitation of an electron from the occupied to the unoccupied states, creating an electron-hole pair, (ii) acceleration of the electron and hole by the external driving field across the Brillouin zone, (iii) recombination of the electron-hole pair, resulting in the emission of high-energy radiation. In our simulations, the electronic bands are half-occupied; thus there are three primary pathways in the ESSH-topological phase: from the valence band to the conduction band (VB-CB transitions), from the valence band to edge states (VB-ESs transitions), and from edge states to the conduction band (ESs-CB transitions). The comparison between ESSH-topological (red curve) and ESSH-trivial (green curve) phases suggests that transitions involving ESs and the other pathways interfere {\it constructively}, leading to stronger intensity for ESSH-topological phase. 

The Rice-Mele model with the extended coupling contains four midgap states (MGSs) whose energy levels are shifted to $\pm U$ from zero. The presence of these MGSs provides additional pathways other than VB-CB transition. For half-occupied states, the additional pathways include  VB-higher MGSs, lower MGSs-higher MGSs, and lower MGSs-CB transitions. Although Rice-Mele model has more pathways than ESSH-topological model, Fig.  \ref{fig:HHG_compare_essh_rice_mele} shows that its HHG intensity (blue curve) is weaker than the topological phase's (red curve) for harmonics less than $20$, corresponding to $E_{gap}/\omega$, suggesting the MGSs are more likely to cause {\it destructive} interference with other transitions. 

\begin{figure}[tbp]
	\centering
	\includegraphics[width=0.97\linewidth]{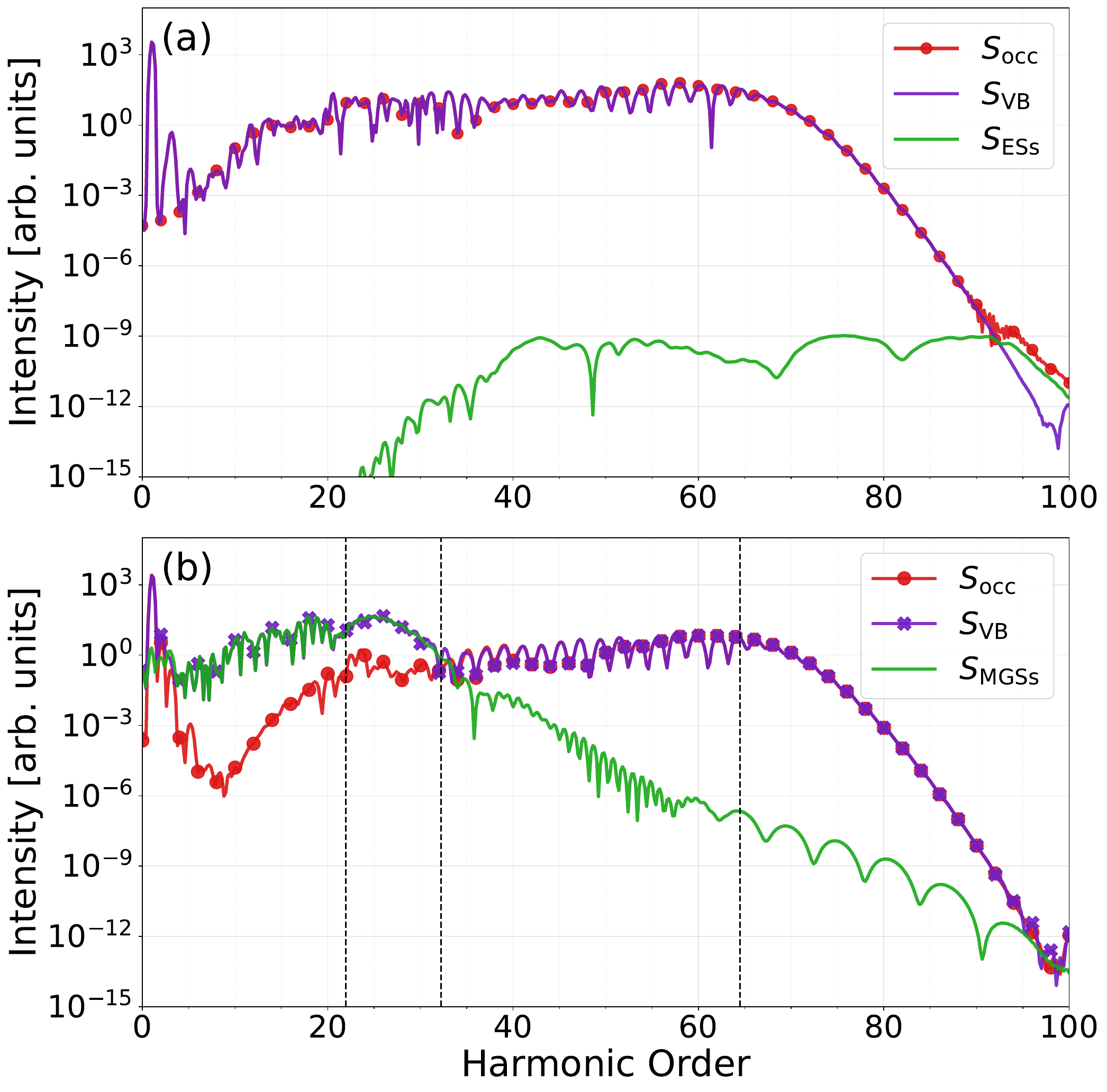}
	\caption{Decomposition analysis of harmonic spectrum comparing the contributions from the total dipole acceleration ($S_{\mathrm{occ}}$), valence-band states ($S_{\mathrm{VB}}$), and edge or midgap states ($S_{\mathrm{ESs}}$ or $S_{\mathrm{MGSs}}$) for (a) ESSH-topological phase, (b) Rice-Mele model. Black vertical dashed lines in the lower panels indicate characteristic harmonic orders corresponding to the bulk bandgap ($E_{\mathrm{gap}}/\omega \approx 22$), one quarter of the maximum CB-VB energy difference ($E_{\mathrm{max}}/4\omega \approx 32$), and one half of the maximum CB-VB energy difference ($E_{\mathrm{max}}/2\omega \approx 64$).}
	\label{fig:hhg_compare_n}
\end{figure}

The decomposed spectrum is shown in Fig. \ref{fig:hhg_compare_n}, which panel (a) shows the results for the ESSH-topological model, agreeing with those reported for conventional SSH with winding number $=1$~\cite{PhysRevA.110.033113}. The contribution from the VB, $S_{\mathrm{VB}}$, dominates and closely matches the full response $S_{\mathrm{occ}}$, indicating that the HHG originates almost entirely from the VB. The edge states contribution $S_{\mathrm{ESs}}$ is almost negligible, owing to the symmetric transition dipole moments (TDMs) that lead to destructive interference between ES-related transitions. In contrast, the Rice-Mele model breaks the symmetry of the TDMs due to the introduction of staggered on-site potentials. As shown in Fig. \ref{fig:hhg_compare_n} (b), in the regime corresponding to harmonic orders in $[0, E_{\mathrm{max}}/4\omega] \approx [0, 32]$, the midgap states contribution $S_{\mathrm{MGSs}}$ becomes comparable to $S_{\mathrm{VB}}$, indicating that the midgap states are active contributors to harmonic generation in this energy range. However, beyond harmonic order $\approx 32$, $S_{\mathrm{MGSs}}$ rapidly diminishes and becomes negligible. Interestingly, we find that the total spectrum $S_{\mathrm{occ}}$ is significantly weaker than either $S_{\mathrm{VB}}$ or $S_{\mathrm{MGSs}}$ alone, particularly in the low-energy regime. Given that $S_{\mathrm{VB}} \approx S_{\mathrm{MGSs}} > S_{\mathrm{occ}}$, this suggests that strong {\it destructive} interference occurs between the VB and MGSs dipole accelerations, suppressing the net harmonic intensity. In the intermediate region $[32, 64]$, the midgap states contribution continues to fade away and therefore the total response becomes dominated by the VB.

\begin{figure}[tbp]
	\centering
	\includegraphics[width=0.98\linewidth]{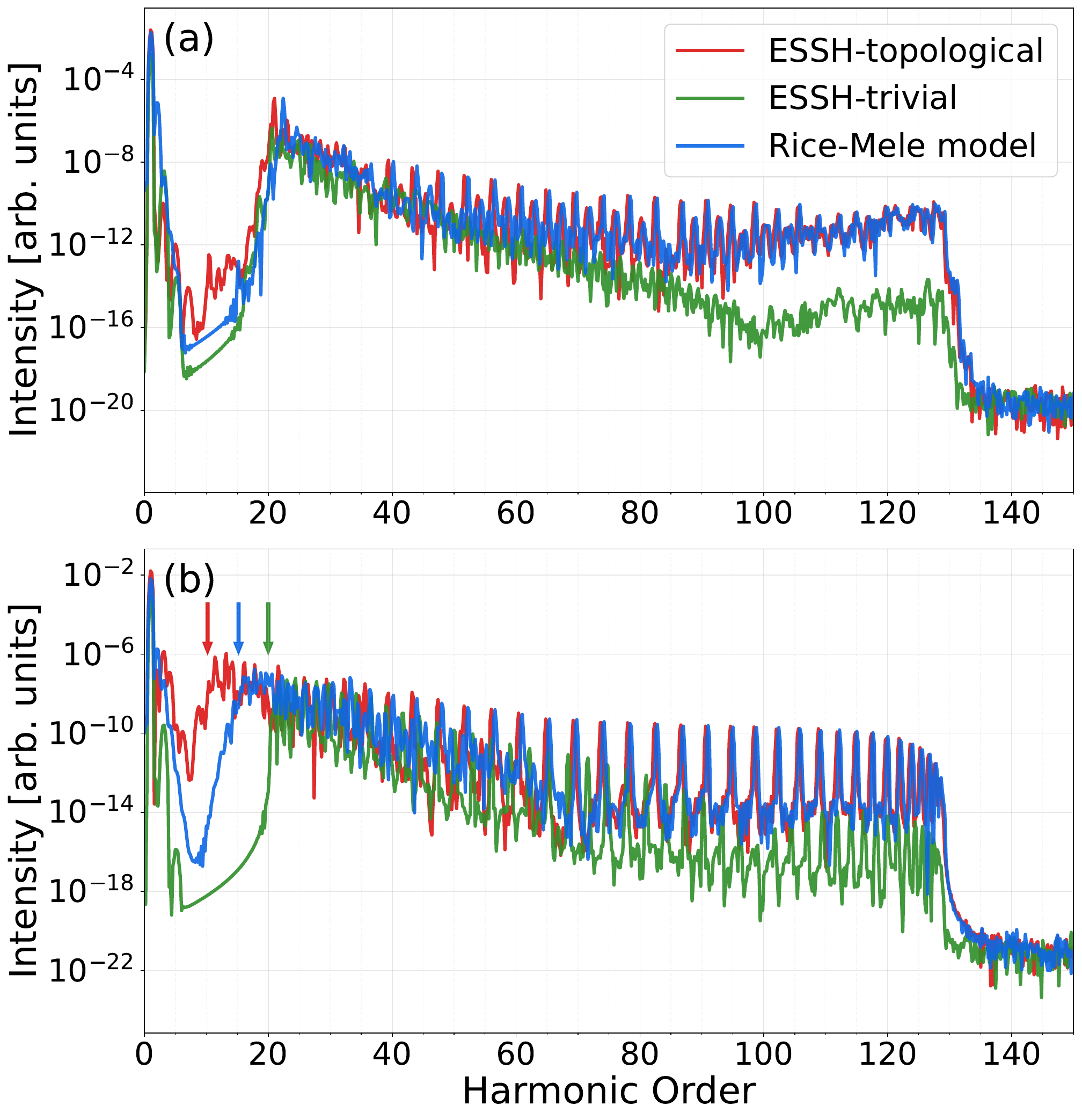}
	\caption{Harmonic spectrum of the ESSH-trivial phase, ESSH-topological phase, and the Rice-Mele model under localized illumination. (a) Illumination centered in the middle of the chain ($x_0=0$). (b) Illumination applied at the boundary ($x_0=95$). Three arrows indicate characteristic harmonic orders for three different models, from left to right: 10 ($E_{\mathrm{gap}}/2$ for ESSH model), 15 ($\Delta E_2$  from occupied MGSs to CB valley for Rice-Mele model), and 20 ($E_{\mathrm{gap}}$ for ESSH model).} 
	\label{fig:HHG_local_light_compare_essh_ricemele}
\end{figure}

Next, we present harmonic spectra of the models under local illumination. Fig.~\ref{fig:HHG_local_light_compare_essh_ricemele} shows the harmonic spectrum of half-occupancy for $N=100$. We localized the pulse either in the center ($x_0 = 0\ \mathrm{a.u.}$, $x_l = 10\ \mathrm{a.u.}$), or at the boundary of the chain ($x_0 = 95\ \mathrm{a.u.}$, $x_l = 10\ \mathrm{a.u.}$), to distinguish the spectral contributions from bulk states and boundary-localized states. We can clearly observe that when the pulse is localized in the center of the chain, as shown in panel (a), all three models, ESSH-trivial, ESSH-topological, and the Rice-Mele model, exhibit qualitatively similar harmonic spectrum throughout all harmonic orders, as the ESs and MGSs are more localized near boundaries.

In contrast, when the pulse is applied at the boundary, as shown in panel (b), significant spectral differences emerge, particularly below the bulk bandgap regime corresponding to harmonic order $\approx 20$. The intensity increases to a local maximum, as indicated by the arrows in the figure, and then decreases. In the ESSH-topological phase, the local maximum appears near harmonic order $10$, corresponding to half of the bulk bandgap, consistent with VB-ESs transitions.
In the ESSH-trivial phase, the local maximum appears at harmonic orders $20$, corresponding to the bulk gap. 
Meanwhile, the Rice-Mele model displays a spectral local maximum at order $15$, indicating transition between occupied MGs and the CB minimum. These observations demonstrate that the spatial profile of the laser pulse and wave functions play crucial roles in shaping the harmonic spectra. 
\begin{figure*}[tbp]
	\centering
	\includegraphics[width=0.955\linewidth]{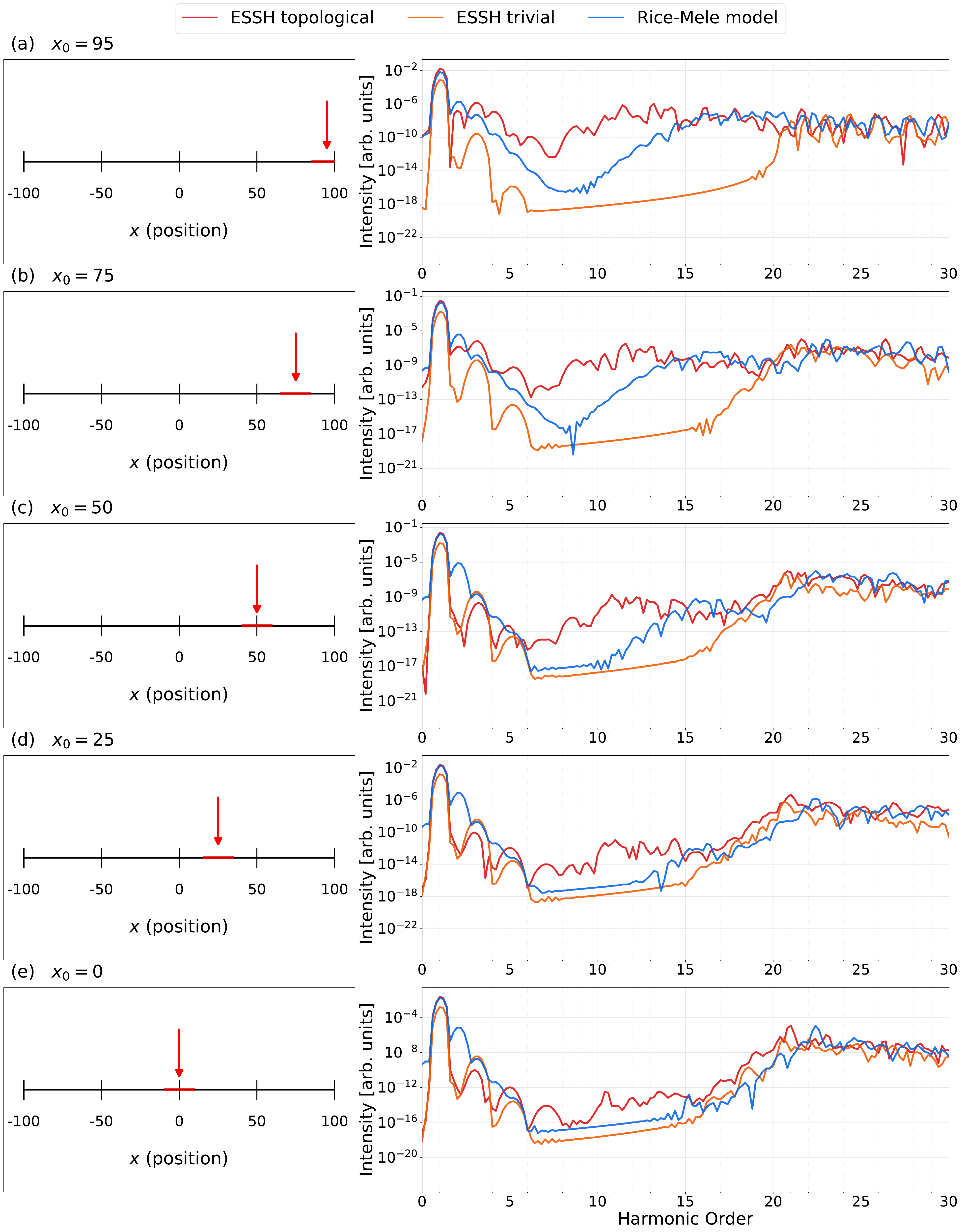}
	\caption{Harmonic spectrum under localized illumination at different positions along the chain with fixed $N=100$ and $x_l=10$. Left panels illustrate the schematic diagram for position of localized pulse (red arrow) and illumination range (red line). Right panels display spectrum for pulses centered at (a) $x_0=95$, (b) $x_0=75$, (c) $x_0=50$, (d) $x_0=25$, and (e) $x_0=0$.}
	\label{fig:HHG_compare_light_positions}
\end{figure*}

We continuously shift the center of the localized pulse along the chain to investigate how the HHG response evolves as the illumination center shifts. The results are shown in Fig.~\ref{fig:HHG_compare_light_positions}. Panels (a)--(e) display results for illumination center from the chain boundary to chain center. The left panels schematically depict the positions of the laser fields. The right panels show the spectrum of each model. 
For the ESSH-topological phase (red curve), even-order harmonics emerge when the illumination center is near the boundary, indicating that local illumination breaks the inversion symmetry of the system. The even-order harmonics diminish as the illumination position shifts toward the chain center, implying that the edge states are responsible for the even-order harmonic emission. 
The spectrum of the ESSH-trivial phase (orange curve) does not show significant differences as the illumination position shifts from the boundary to the bulk, consistent with the delocalized nature of all bulk states. 
In panel (e), the ESSH-topological and trivial overlap almost entirely. This behavior can be attributed to the localization of ESs, such that the illumination at the chain center does not excite the ESs.
The Rice-Mele model (blue curve) is distinguished by the presence of even-order harmonics below harmonic order 6, irrespective of the illumination position.


\section{\label{sec: local_illuminating} Symmetry origin of even harmonics under local illumination}

\begin{figure*}[htbp]
	\centering
	\includegraphics[width=0.9\textwidth]{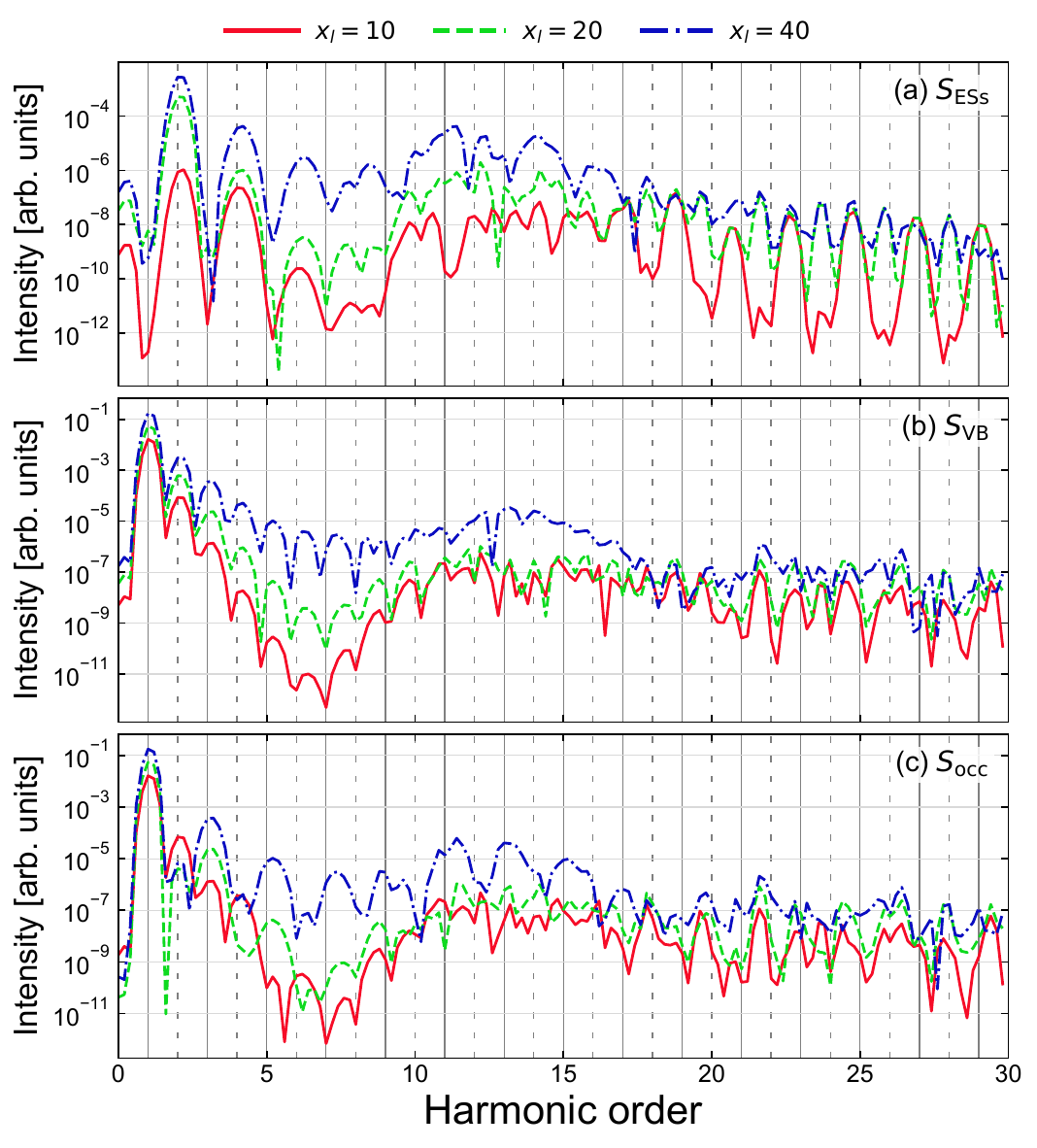}
	\caption{Decomposition analysis for ESSH-topological phase under local illumination with various $x_l$ with fixed $x_0=N-5$ for $N=100$.}
	\label{fig:local_decomp}
\end{figure*}

To understand the intriguing even harmonics emerged in the ESSH-topological phase under local illumination on the boundary, we performed the decomposition analysis. The results are given in Fig. \ref{fig:local_decomp}. As shown in panel (a), even-order harmonics are dominant in $S_{\rm ESs}$, while, in panel (b), even and odd-order harmonics are both present in $S_{\rm VB}$. As shown in panel (c), the even-order harmonics are suppressed in $S_{\rm occ}$ for 
$x_\ell = 20$ and $40$, indicating that the even-order 
contributions from the edge state and valence band interfere 
destructively in the total occupied-state response. For smaller $x_\ell$, however, even-order harmonics remain visible in $S_{\rm occ}$. This is because the edge state decays exponentially into the bulk, and a smaller illuminated window does not capture the full weight of the edge state. As a result, the even-order contribution from $S_{\rm ES}$ is insufficient to fully cancel that from $S_{\rm VB}$, leaving a residual even-order signal in $S_{\rm occ}$. 



Additionally, we have calculated the harmonic spectrum and the decomposition analysis for the topological SSH model with different illumination positions. The results are given in Appendix \ref{app:sshlocal}. The decomposition shows that even-order harmonics dominate in $S_{\rm ES}$, consistent with the results for ESSH-topological phase. 
 
We identify the even harmonics in $S_{\rm ESs}$ as a result of the zero-energy character of the edge states and particle-hole symmetry. Local illumination breaks inversion symmetry and thereby allows even-order harmonics, while the zero-energy edge states together with particle-hole symmetry suppress odd field orders in the edge-state response. The analysis is given below. Starting from the Schr\"{o}dinger equation $i\partial_t|\psi_{E}(t)\rangle=(H_0+E(t)D)|\psi_{E}(t)\rangle$, where $H_0$ is the Hamiltonian of the system that preserves $\Gamma$ and $K$, the initial state $|\psi_{E}(0)\rangle$ is the eigenstate of $H_0$ and $D$ is the localized position operator, written as $D=\sum_{j}d_j\hat{c}_j^{\dagger}\hat{c}_j$, where $d_j$ is defined in Eq. \eqref{eq:localpot}, for $|x_j-x_0|<x_{\ell}$ and $0$ elsewhere. After the particle-hole symmetry transformation, the full Hamiltonian becomes $\mathcal{P}(H_0+ED)\mathcal{P}^{-1}=-H_0+ED$ and the Schr\"{o}dinger equation becomes
\begin{equation}
	i\partial_t|\mathcal{P}\psi_{E}(t)\rangle =(H_0-E(t)D)\mathcal{P}|\psi_{E}(t)\rangle,
\end{equation}
which shows that with the initial condition $\mathcal{P}|\psi_E(0)\rangle$,  $\mathcal{P}|\psi_{E}(t)\rangle$ is the solution for a reversed electric field, denoted by $|\psi_{-E}(t)\rangle$

We denote the expectation value of $\hat{x}$ for $|\psi_{E}(t)\rangle$ with initial condition $|\psi_{E}(0)\rangle$ as
\begin{equation}
	X_{\psi_0}[E](t)\equiv\langle\psi_{E}(t)|\hat{x}|\psi_{E}(t)\rangle
\end{equation}
and that for 
$|\psi_{-E}(t)\rangle$ with initial condition $\mathcal{P}|\psi_{E}(0)\rangle$ as
\begin{equation}
	X_{\mathcal{P}\psi_0}[-E](t)\equiv\langle\psi_{-E}(t)|\hat{x}|\psi_{-E}(t)\rangle. 
\end{equation}
Substituting $|\psi_{-E}(t)\rangle$ with $ \mathcal{P}|\psi_{E}(t)\rangle$ and using the antiunitary property of $\mathcal{P}$, we obtain
\begin{eqnarray}
X_{\mathcal{P}\psi_0}[-E](t)
&=&\langle\mathcal{P}\psi_{E}(t)|\hat{x}|\mathcal{P}\psi_{E}(t)\rangle\nonumber\\
&=&\langle\psi_E(t)|\mathcal{P}^{-1}\hat{x}\mathcal{P}|\psi_E(t)\rangle^*\nonumber\\
&=&\langle\psi_E(t)|\mathcal{P}^{-1}\hat{x}\mathcal{P}|\psi_E(t)\rangle\nonumber\\
&=&\langle\psi_{E}(t)|\hat{x}|\psi_{E}(t)\rangle\nonumber\\
&=&X_{\psi_0}[E](t), 
\end{eqnarray}
where the second line applies the antiunitary inner product identity $\langle\mathcal{P}\phi|\hat{O}|\mathcal{P}\psi\rangle = 
\langle\phi|\mathcal{P}^{-1}\hat{O}\mathcal{P}|\psi\rangle^*$, the third line uses $\mathcal{P}^{-1}\hat{x}\mathcal{P} = \hat{x}$, and the complex conjugate drops because $\hat{x}$ is a Hermitian operator. 

Next, we consider the zero-energy subspace of $H_0$. 
If $H_0|\psi_n\rangle=\epsilon_n|\psi_n\rangle$, then the particle-hole symmetry implies 
\begin{equation}
	H_0\mathcal{P}|\psi_n\rangle=-\epsilon_n\mathcal{P}|\psi_n\rangle
\end{equation}
and for the zero eigenenergy state $\mathcal{P}|\psi_0\rangle\propto |\psi_0\rangle$ \footnote{This holds true because the edge zero modes localize on one of the sublattices. The left (right) edge states live entirely on the A (B) sublattice and can be chosen as particle-hole-invariant states. For finite chains, despite hybridizations between left and right edge zero modes, local illumination predominantly couples to the edge state near the illuminated boundary}, leading to $X_{\mathcal{P}\psi_0}[-E](t)=X_{\psi_0}[-E](t)$. Therefore, the time-dependent expectation values of the position operator for the initial state being the zero-mode is
\begin{equation}
	X_{0}[E](t)=X_{0}[-E](t).
	\label{eq:zero_evenparity}
\end{equation}
This is in sharp contrast to the case where the initial state is a valence (conduction) band state $|\psi_v\rangle$ at energy $-\epsilon \neq 0$. For the valence band, $\mathcal{P}|\psi_v\rangle \propto |\psi_c\rangle$, the symmetry partner is the conduction band state at $+\epsilon$. Therefore, $X_{\psi_v}[E](t)=X_{\mathcal{P}\psi_v}[-E](t) = X_{\psi_c}[-E](t) \neq X_{\psi_v}[-E](t)$ in general. The two sides of the general identity $X_{\psi_0}[E](t) = X_{\mathcal{P}\psi_0}[-E](t)$ then involve \emph{different} initial states, and no field-parity constraint on $X_{\psi_v}[E](t)$ follows. 

Finally,  the connection between Eq.~\eqref{eq:zero_evenparity} and the even harmonic structure of the spectrum is established through time-dependent perturbation theory. The derivation can be found in Appendix~\ref{app:tdep}.  It can be shown that $X_M[E](t)$, where $M$ labels the eigenstate at $t=0$, can be written as sum of the series expansion in powers of driving field, 
\begin{eqnarray}
	X_M[E](t)=\sum_{\ell=0}^{\infty}X_M^{\ell}[E](t),
\end{eqnarray}
where $X_M^{\ell}[E](t)$ is the contribution of order $\ell$ in field amplitude and satisfies 
\begin{equation}
	X_M^{\ell}[E](t)=(-1)^{\ell}X_M^{\ell}[-E](t)
	\label{eq:minusfactor}
\end{equation}
when reversing the electric field. 
Combining with \eqref{eq:zero_evenparity}, which must be satisfied order by order in perturbation theory for arbitrary field amplitude, we have for edge zero mode $X_{0}^{\ell}[E](t)=(-1)^{\ell}X_{0}^{\ell}[E](t)$. Thus, $X_{0}^{\ell}[E](t)$ vanishes for odd $\ell$, leaving strong even harmonics in $S_{\rm ES}$. 
For the ESSH model with higher winding number, each edge hosts more than one zero mode. Within the zero-energy subspace, one can always choose a particle-hole-invariant basis satisfying $\mathcal{P}|\psi_0\rangle\propto|\psi_0\rangle$, so Eq. \eqref{eq:zero_evenparity} extends to each zero mode.

\section{\label{sec: summary} Summary} 
In conclusion, we investigate high-order harmonic generation in the extended Su-Schrieffer-Heeger (ESSH) and the Rice-Mele models. The roles of edge (ESs) and midgap states (MGSs) are studied via decomposition analysis. 
Under global illumination, for the ESSH model, we find that the intensity in topological phase is stronger than that in the trivial phase. For the Rice-Mele model, the staggered on-site potential splits the degenerate zero modes into two pairs of midgap states (MGSs). We find that these MGSs do not enhance but instead suppress the overall harmonic intensity, particularly below the bulk bandgap regime. Furthermore, the decomposition analysis shows that in the topological phase of the ESSH model, the valence band (VB) contribution dominates the spectral intensity, whereas the contribution from the edge states is negligible. By contrast, in the Rice-Mele model, the MGSs contribution becomes comparable to that of the VB at low harmonic orders but interferes destructively with the VB contribution, yielding a net response much weaker than either individual contribution.

Local illumination distinguishes the spectrum shaped by edge states, particularly for harmonic orders $<6$. For the ESSH-topological phase, when the illumination center is on the boundary, the even-order harmonics become significant, indicating inversion symmetry breaking. The decomposition analysis shows that the edge states give rise to dominant even-order harmonics and the valence bands contribute to even- and odd-order harmonics. The full spectrum is a result of the destructive interference between the contributions from valence band and edge states. The incomplete destructive interference leads to remnant even-order harmonics in the full spectrum when edge states are illuminated partially. For ESSH-trivial phase and Rice-Mele model, the change of illumination position does not vary the HHG spectrum significantly. We identify the origin of the even-order harmonics as the zero-energy character of the edge states and the particle-hole symmetry of the system. This phenomenon also holds for the SSH model, as confirmed by numerical calculations.

\acknowledgments
The authors acknowledge Dr. Chuan Yu's insightful advice and Dr. Xiao Zhang's helpful suggestions. C.-T.L. and H.-C.H. acknowledge the support from the National Science and Technology Council (NSTC) under Grant No. 113-2628-M-004-001-MY3 and the National Center for Theoretical Sciences (NCTS) in Taiwan. J.-S.Y. acknowledges support from the National Science and Technology Council (NSTC), Taiwan, under Grant No. NSTC 113-2112-M-003-015 and No. NSTC 114-2112-M-003-005, from “Higher Education Sprout Project“ of National Taiwan Normal University and the Ministry of Education (MOE), Taiwan, and from TG 3.2 of NCTS.

The authors used Claude Sonnet 4.6 (Anthropic) and ChatGPT (GPT-5.6 Sol, OpenAI) to assist with the perturbative expansion and formulation of symmetry analysis. The AI-assisted analysis was guided by the authors, who independently checked the derivations and critically reviewed and revised the resulting content. The authors take full responsibility for the accuracy of the manuscript.
\clearpage
\appendix
\section{Su-Schrieffer-Heeger model under local illumination}\label{app:sshlocal}
\begin{figure*}[htbp]
	\centering	\includegraphics[width=0.95\linewidth]{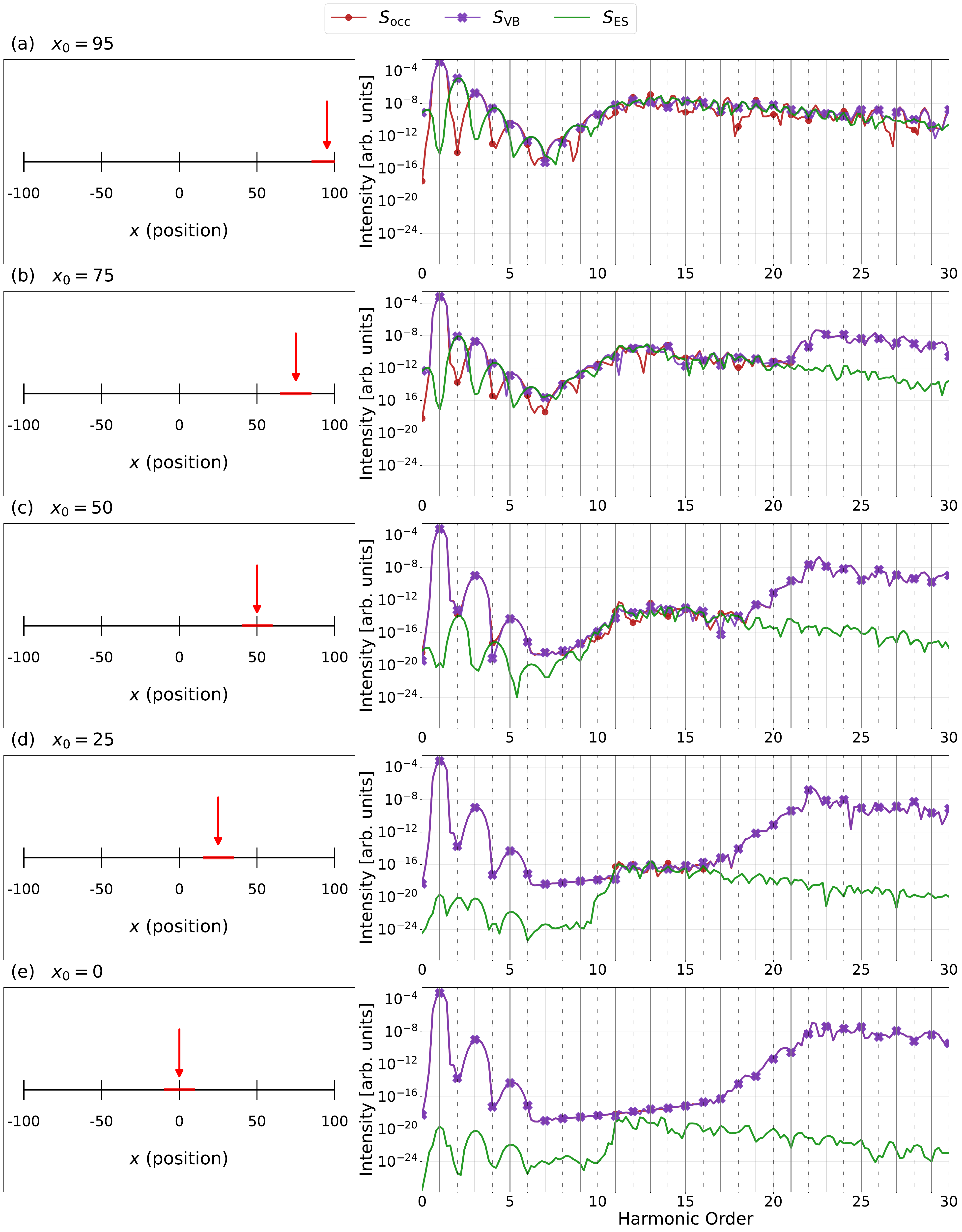}
	\caption{Decomposition analysis of the SSH model in the topological phase, comparing $S_{\mathrm{occ}}$, $S_{\mathrm{ES}}$, and $S_{\mathrm{VB}}$ for chain length $N=100$ under local illumination. The right panels in (a-e) show the harmonic spectrum for different illumination center ($x_0$), as schematically illustrated in the left panels. The vertical dashed (solid) lines indicate even (odd) harmonic orders. The model parameters for the calculation are $v=-0.10, w=-0.18$, and $\delta=-0.15$.}
		\label{fig:SSH_LL_CD_position_comparison}
	\end{figure*}

We supplement the harmonic spectra for the SSH model under local illumination. In Fig.~\ref{fig:SSH_LL_CD_position_comparison}, 
$S_{\mathrm{occ}}$ is the full (undecomposed) spectrum for half-occupancy, $S_{\mathrm{VB}}$ represents the contribution from the valence bands, and $S_{\mathrm{ESs}}$ represents the contribution from the occupied edge states. From panels~(a)--(e), the center of the laser pulse shifts from the right boundary to the center of the chain. Our $S_{\mathrm{occ}}$ agrees with that in \cite{PhysRevB.99.195428}, in which the full spectrum is qualitatively similar to that obtained under global illumination, irrespective of the local illumination position. However, the decomposition analysis reveals that $S_{\mathrm{ES}}$ and $S_{\mathrm{VB}}$ change drastically as the illumination position shifts.

When the illumination position is near the boundary, as shown in panel~(a), $S_{\mathrm{ES}}$ (green curve) exhibits dominant even-order harmonics, whereas $S_{\mathrm{VB}}$ (purple curve) contains both even- and odd-order harmonics. The even-order 
harmonics in $S_{\mathrm{VB}}$ and $S_{\mathrm{ES}}$ interfere destructively, leading to vanishing even-order harmonics in 
$S_{\mathrm{occ}}$.

As the illumination position moves toward the center of the chain, as shown in panels~(b)--(e), the intensity of even-order harmonics decreases in both $S_{\mathrm{ES}}$ and $S_{\mathrm{VB}}$. In panel~(e), inversion symmetry is restored as the illumination is centered on the chain, and all spectra display dominant odd-order harmonics. The intensity of $S_{\mathrm{ES}}$ becomes much weaker than that of $S_{\mathrm{VB}}$.

\section{Perturbative expansion and field reversal parity}\label{app:tdep}
For the unperturbed Hamiltonian $H_0$, the eigenequation is written as  $H_0|\psi_n\rangle=\epsilon_n|\psi_n\rangle$. The perturbing term is written as $H'=E(t)D$ and turned on at $t=0$. 
The time-dependent quantum state is 
\begin{equation}
	|\Psi_M(t)\rangle=\sum_j|\psi_j\rangle c_{Mj}(t)e^{-i\epsilon_jt},
\end{equation} 
where $M$ denotes the index of the unperturbed eigenstate before $t=0$. The time-dependent coefficients $c_{Mj}(t)$ satisfy the first-order differential equation
\begin{equation}
	\partial_t{c_{Mi}(t)}=-i\sum_jE(t)D_{ij}c_{Mj}(t)e^{-i\Delta_{ji}t},
\end{equation}
where $D_{ij}\equiv\langle\psi_i|D|\psi_j\rangle$ and $\Delta_{ji}\equiv\epsilon_j-\epsilon_i$.

The coefficients can be solved iteratively by substituting the lower order solution into the integral, 
\begin{eqnarray}
	c_{Mi}^{(\ell+1)}(t)=-i\int_0^t dt'E(t')\sum_jD_{ij} c^{(\ell)}_{Mj}(t')e^{-i\Delta_{ji}t'},
\end{eqnarray}
where $\ell=0,1,2,3...$ denotes the order of expansion series. By substituting the initial condition $c_{Mi}^{(0)}=\delta_{Mi}$, it is straightforwardly seen that $c_{Mi}^{(\ell)}(t)$ is proportional to the $\ell$-th power of the field amplitude. The expectation value of the position operator is given by the summation of all orders
\begin{equation}
	X_M(t)\equiv\langle\Psi_M(t)|\hat{x}|\Psi_M(t)\rangle=\sum_{\ell=0}^{\infty}X_M^{(\ell)}(t)
\end{equation}
where 
\begin{eqnarray}
	X^{(\ell)}_M(t)=\sum_{p+q=\ell}\sum_{i,j}c^{(p)*}_{Mi}(t)x_{ij}c^{(q)}_{Mj}(t)e^{-i\Delta_{ji} t}
\end{eqnarray}
and $x_{ij}\equiv\langle\psi_i|\hat{x}|\psi_j\rangle$. 
$X^{(\ell)}_M(t)$ is proportional to the $\ell$-th power of the field amplitude. When electric field reverses sign, an additional $(-1)^{\ell}$ factor arises, as written on the right hand side of Eq. \eqref{eq:minusfactor}. 

\clearpage
\bibliography{apssamp}
\end{document}